\documentclass{aa}  

\usepackage{float}
\usepackage{graphicx}
\usepackage{txfonts}
\usepackage{array}
\usepackage{natbib}
\bibpunct{(}{)}{;}{a}{}{,}   
\usepackage{hyperref}
\def\kms{km\,s$^{-1}$}
\begin{document}
  
   \title{The Galactic evolution of sulphur as traced by globular clusters
  \thanks{Based on observations made with ESO telescopes at the La Silla Paranal Observatory under programmes ID 091.B-0171(A).}
  }
   \titlerunning{Sulphur in GCs}
   
    \author{N. Kacharov\inst{1}\thanks{Member of the International Max Planck Research School for Astronomy and Cosmic Physics at the University of Heidelberg, IMPRS-HD, Germany.}
          \and
            A.~Koch\inst{1}
          \and
            E.~Caffau\inst{2}
          \and
            L.~Sbordone\inst{3,4}
          }

   \institute{Landessternwarte, Zentrum f\"{u}r Astronomie der Universit\"{a}t Heidelberg, K\"{o}nigstuhl 12, D-69117 Heidelberg, Germany\\
              \email{n.kacharov@lsw.uni-heidelberg.de}
         \and
              GEPI, Observatoire de Paris, CNRS, Univ. Paris Diderot, Place Jules Janssen, 92195, Meudon, France
         \and
              Millenium Institute for Astrophysics, Av. Vicu\~{n}a Mackenna 4860, 782-0436 Macul, Santiago, Chile
         \and
              Pontifica Universidad Cat\'{o}lica de Chile, Av. Vicu\~{n}a Mackenna 4860, 782-0436 4860, Macul, Santiago, Chile
              }

   \date{Received 26 Sep. 2014; accepted 9 Mar. 2015}


  \abstract
   {Sulphur is an important, volatile $\alpha$ element but its role in the Galactic chemical evolution is still uncertain and more observations constraining the sulphur abundance in stellar photospheres are required.}
   {In this work we derive the sulphur abundances in red giant branch (RGB) stars in three Galactic halo globular clusters (GC) that cover a wide metallicity range ($-2.3<$[Fe/H]$<-1.2$), namely M\,4 (NGC\,6121), M\,22 (NGC\,6656), and M\,30 (NGC\,7099). The halo field stars show a large scatter in the [S/Fe] ratio in this metallicity span, which is inconsistent with canonical chemical evolution models. To date, very few measurements of [S/Fe] exist for stars in GCs, which are good tracers of the chemical enrichment of their environment. However, some light and $\alpha$ elements show star-to-star variations within individual GCs and it is yet unclear whether the $\alpha$ element sulphur also varies between GC stars.}
   {We used the the infrared spectrograph CRIRES to obtain high-resolution (R~$\sim50000$), high signal-to-noise (SNR~$\sim200$~per px) spectra in the region of the S I multiplet 3 at $1045$\,nm for $15$ GC stars selected from the literature ($6$ stars in M\,4, $6$ stars in M\,22 and $3$ stars in M\,30). Multiplet 3 is better suited for S abundance derivation than the more commonly used lines of multiplet 1 at $920$\,nm, since its lines are not blended by telluric absorption or other stellar features at low metallicity.}
   {We used spectral synthesis to derive the [S/Fe] ratio of the stars assuming local thermodynamic equilibrium (LTE). We find mean [S/Fe]$_{\mathrm{LTE}}=0.58\pm0.01\pm0.20$\,dex (statistical and systematic error) for M\,4, [S/Fe]$_{\mathrm{LTE}}=0.57\pm0.01\pm0.19$\,dex for M\,22, and [S/Fe]$_{\mathrm{LTE}}=0.55\pm0.02\pm0.16$\,dex for M\,30. The negative NLTE corrections are estimated to be in the order of the systematic uncertainties. We do not detect star-to-star variations of the S abundance in any of the observed GCs with the possible exception of two individual stars, one in M\,22 and one in M\,30, which appear to be highly enriched in S.}
   {With the tentative exception of two stars with measured high S abundances, we conclude that sulphur behaves like a typical $\alpha$ element in the studied Galactic GCs, showing enhanced abundances with respect to the solar value at metallicities below [Fe/H]~$-1.0$\,dex without a considerable spread.}

   \keywords{Stars: abundances --
             Globular clusters: general --
             Globular clusters: individual: M\,4, M\,22, M\,30 --
             Galaxy: halo --
               }

   \maketitle
%

\section{Introduction}

Sulphur is one of the less studied $\alpha$-elements (produced by sequential capturing of He nuclei) in stars.
In general, the production site of the $\alpha$-elements such as O, Ne, Mg, Si, S, Ar, Ca, and Ti is mainly associated with the eruptions of type II supernovae (SNe II). In contrast, the major fraction of the iron-peak elements abundance comes from SNe Ia, which contribute little or no $\alpha$-elements.
The different timescales of the occurrence of SNe Ia or II makes the [$\alpha$/Fe] ratio a powerful tool for diagnosing the chemical evolution and star formation history (SFH) of any stellar population \citep{tinsley1979,matteucci+brocato1990,mcwilliam+2013}.
In the Milky Way, the metal poor stars ([Fe/H]~$<-1.0$\,dex) form a plateau at [$\alpha$/Fe]~$\sim0.4$\,dex. With the onset of SNe Ia, at metallicities above [Fe/H]~$\sim-1.0$\,dex, the [$\alpha$/Fe] ratio starts to gradually drop reaching solar values \citep{mcwilliam1997}.
In dwarf galaxies that experienced slower star formation, lower values of [$\alpha/$Fe] are observed even at low metallicities, owing to the lower enrichment rate of the interstellar medium from SNe II \citep[see][and references therein]{shetrone+2001,shetrone+2003,venn+2004,tolstoy+09,hendricks+2014}.
Although different $\alpha$ elements are produced on similar timescales, they show element-to-element scatter due to different production mechanisms, either through He burning in the cores of massive stars, or during the SNe explosions themselves.
In particular, sulphur (along with Si, Ar, and Ca) is produced during both the O-shell burning and explosive oxygen burning phases \citep{limongi+chieffi2003}.

Sulphur is an especially interesting $\alpha$-element because it is not depleted onto dust \citep{ueda+2005} and is thus a genuine tracer of the "true" S abundance of the ISM and the stellar photospheres.
In fact, sulphur is a preferred tracer of the $\alpha$ abundance in interstellar gas \citep{garnett1989,savage+sembach1996} and in high redshift star forming environments such as damped Ly-$\alpha$ systems \citep{centurion+2000,nissen+2004,bowen+2005}.
The existing sulphur measurements in halo field stars show that the [S/Fe] ratio is increasing with decreasing [Fe/H] at higher metallicities, reaching an average value of about $0.4$\,dex but its behaviour at lower metallicities is still rather controversial. Different studies report either a bimodality of the [S/Fe] ratio with a flat plateau and a branch of increasing abundance, reaching [S/Fe]~$\sim0.8$\,dex, or a plateau with a large scatter \citep[see e.g.][and references therein]{nissen+2004,caffau+2005,nissen+2007,spite+2011,jonsson+2011,matrozis+2013}.   

Globular clusters (GCs) represent the oldest stellar populations in the Milky Way and are examples of very dense environments.
Long time considered as simple stellar populations, nowadays we recognize their complex SFH through precise abundance analysis of a variety of chemical elements in individual cluster member stars. All GCs, studied to date, present significant spreads and certain unique correlations in their light and $\alpha$ element abundances -- generally taken as a clue for multiple enrichment phases \citep[see the recent review by][]{gratton+2012}.
Still, sulphur abundances have been derived in very few GCs due to the difficulties that such measurements entail. Sulphur lines in the optical spectral range are generally very weak in giant stars. At the same time, these brighter stars are often the only accessible tracers in remote and faint GCs. To date, the strong multiplet 1 ($920$\,nm) has been used to obtain sulphur abundances in GCs \citep{caffau+2005b,sbordone+2009,koch+caffau2011,caffau+2014} but these lines are often blended with telluric absorption lines and one of them is positioned in the blue wing of the Paschen $\zeta$ line and thus cannot always be reliably measured.
This makes the Mult. 3 at $1045$\,nm particularly suited to measure the sulphur abundance. Even if the lines are not as strong as the components of Mult. 1, a big advantage is that there are no telluric features in this wavelength range. Observing the $1045$\,nm sulphur lines leads to reliable results even at lower metallicities and will give independent and possibly more accurate sulphur abundances in GCs than using Mult. 1 \citep{caffau+2007,caffau+2010,jonsson+2011}.

The metal rich GC Ter\,7 ([Fe/H]~$=-0.32$\,dex) is the first GC, for which S abundances were ever derived. The three stars measured by \citet{caffau+2005b} are also the only extragalactic stars with known [S/Fe] ratio, since Ter\,7 is associated with the disrupted Sagittarius dwarf spheroidal galaxy. \citet{caffau+2005b} determined mean S abundance in Ter\,7 slightly lower (by $\sim0.1$\,dex) than in Galactic stars at similar metallicity.
\citet{sbordone+2009} analysed $9$ stars in the GC 47\,Tuc ([Fe/H]~$=-0.72$\,dex) and $4$ stars in NGC\,6752 ([Fe/H]~$=-1.54$\,dex), deriving [S/Fe]~$=0.18\pm0.14$ and [S/Fe]~$=0.49\pm0.15$\,dex for both GCs, respectively, consistent with the results of field stars.
Curiously, \citet{sbordone+2009} reported a star-to-star variation in the observed [S/Fe] ratio in 47\,Tuc, which although not significant with respect to the large measurement uncertainties, showed a strikingly clear correlation with the Na abundance. The authors considered different possibilities for the occurrence of such correlation including that S may be involved in the self-enrichment processes of GCs through the $\mathrm{^{31}P(p,\gamma)^{32}S}$ proton capture reaction. \citet{koch+caffau2011} analysed a single star in the metal poor GC NGC\,6397 ([Fe/H]~$=-2.0$\,dex), determining [S/Fe]~$=0.52\pm0.2$\,dex, consistent with the halo filed population.
We also note the recent work by \citet{caffau+2014}, which provides [S/Fe] measurements for several metal rich open clusters and includes the GC M\,4, analysed in the present work.

Here, we present the homogeneous analysis of $15$ bright RGB stars in three different GCs that cover together a wide range of metallicities ($-2.3<$~[Fe/H]~$<-1.2$\,dex). Such strategy can provide answers to several basic questions. It is interesting to find out whether a dichotomy or a large spread of the [S/Fe] ratio, as observed in the halo field stars, could also be found in GCs.
The generally large uncertainties of the [S/Fe] measurements in the halo field stars are mostly driven by uncertainties of the atmospheric parameters \citep{matrozis+2013} and thus, tracing the [S/Fe] ratio with GCs will lead to more accurate results due to the much better constrained distances, gravities, and effective temperatures of GC stars from colour-magnitude diagrams (CMD).
Additionally, our data will help to check if there are star-to-star variations of the S abundance in the different GCs and whether they correlate with the other $\alpha$ and light element abundances, such as O, Na, Si, Mg, and Al.

The three selected clusters are amongst the closest to the Sun that cover an optimally broad metallicity range and are all well studied objects. The proximity and simple CMDs of M\,4 and M\,30 make them desirable candidates for many observational and theoretical studies testing the current stellar evolutionary models. M\,30 is also notable for its very low metallicity [Fe/H]$=-2.3$\,dex. M\,22, on the other hand, is amongst the most massive GCs. It has a very complex CMD with prominent multiple populations and a large metallicity spread \citep[$-1.9<\rm{[Fe/H]}<-1.6$~dex][]{dacosta+2009,marino+2009}. It is often considered as a remainder of a disrupted dwarf galaxy and will thus add another, important comparison object to our study.

In this paper we present the largest compilation of sulphur abundance measurements in Galactic clusters to date.


\section{Target selection and observations}

We chose our targets from existing high-resolution surveys of these GCs in the visible band; M\,4 stars were chosen from \citet{marino+2008}; M\,22 stars from \citet{marino+2009,marino+2011b}; M\,30 stars from \citet{carretta+2009c,carretta+2009b}.
Our intention was to select bright RGB stars that belong to different populations in the selected GCs. To this end, we targeted stars with different Na and O abundances in M\,4 and M\,30. M\,22, on the other hand, is known to harbour stars with a wide metallicity spread, so we selected stars with different [Fe/H].
The adopted stellar parameters of the observed stars from the literature are presented in Table~\ref{tab:targets}.
Choosing stars for which high-resolution spectroscopy was already available is also important for constraining their atmospheric parameters, such as effective temperature, gravity, microturbulence velocity, and metallicity, necessary for abundance determination. This is only possible if a large set of Fe I and Fe II lines are available, which is not the case with the narrow wavelength coverage of our spectra.
Note, however, that the effective temperatures and gravities of the three stars in M\,30 were determined photometrically by \citet{carretta+2009b} and likely have larger uncertainties. We still opted to rely on them since these parameters were used by \citet{carretta+2009b} to derive abundances of other chemical elements in these stars, with which we aim to compare our S abundance estimates.

To compare both approaches, we compared the stellar parameters for $\sim80$ stars in common between the \citet{marino+2008} and \citet{carretta+2009b} studies.
The effective temperatures determined by \citet{carretta+2009b} are by $\sim100$\,K lower on average than those in \citet{marino+2008} for M\,4 with $\sigma=45$\,K and the $\log g$ values used by \citet{carretta+2009b} are by $0.3$ dex lower on average with $\sigma=0.15$\,dex.
\citet{marino+2008} find a lower spread in microturbulence and a larger spread in metallicities among the common stars, while \citet{carretta+2009b} find the opposite, i.e., a larger spread in microturbulence and a lower spread in metallicity. The average differences are $0.13$\,\kms~ in microturbulence and $0.13$\,dex in [Fe/H] between the two studies.

\begin{table*}
\begin{center}
\caption{Atmospheric parameters and [S/Fe] ratios of the targeted stars.}\label{tab:targets}
{\footnotesize
 \begin{tabular}{ccccccccc>{\bfseries}c>{\bfseries}c}
\hline
 ID & T$_{eff}$ & log g & v$_{mic}$ &  [Fe/H] & [Na/Fe] & [O/Fe] & [Mg/Fe] & [Ca/Fe] & v$_{mac}$\tablefootmark{4} & [S/Fe]$_{\mathrm{LTE}}$\tablefootmark{5} \\
     &    [K]         & [dex] &  \kms        &  [dex]   & [dex]     & [dex]   & [dex]      & [dex]     &  \kms                                     & [dex] \\
\hline
\multicolumn{10}{c}{{\bf M\,4}\tablefootmark{1}} \\
19925 & 4050 & 1.20 & 1.67 & $-$1.02 & 0.51$\pm$0.04 & 0.28$\pm$0.04 & 0.43$\pm$0.06 & 0.19$\pm$0.03 & 6.5$\pm$0.1 & 0.57$\pm$0.02 \\
21191 & 4270 & 1.60 & 1.60 & $-$1.06 & 0.51$\pm$0.04 & 0.34$\pm$0.04 & 0.55$\pm$0.06 & 0.24$\pm$0.03 & 4.8$\pm$0.1 & 0.58$\pm$0.02 \\
27448 & 4310 & 1.57 & 1.58 & $-$1.12 & 0.11$\pm$0.04 & 0.51$\pm$0.04 & 0.50$\pm$0.06 & 0.29$\pm$0.03 & 4.9$\pm$0.1 & 0.57$\pm$0.01 \\
28103 & 3860 & 0.50 & 1.62 & $-$1.08 & 0.17$\pm$0.04 & 0.50$\pm$0.04 & 0.41$\pm$0.06 & 0.16$\pm$0.03 & 6.3$\pm$0.1 & 0.65$\pm$0.02 \\
34006 & 4320 & 1.67 & 1.61 & $-$1.06 & 0.44$\pm$0.04 & 0.25$\pm$0.04 & 0.52$\pm$0.06 & 0.26$\pm$0.03 & 5.7$\pm$0.1 & 0.50$\pm$0.02 \\
36215 & 4300 & 1.59 & 1.53 & $-$1.11 & 0.18$\pm$0.04 & 0.48$\pm$0.04 & 0.52$\pm$0.06 & 0.26$\pm$0.03 & 5.5$\pm$0.2 & 0.59$\pm$0.02 \\
\multicolumn{10}{c}{{\bf M\,22}\tablefootmark{2} } \\
200005 & 4000 & 0.05 & 2.02 & $-$1.94 &$-$0.02$\pm$0.03 &   0.40$\pm$0.04 & 0.46$\pm$0.04 & 0.24$\pm$0.02 & 11.9$\pm$0.2 & 0.78$\pm$0.01 \\
200025 & 4100 & 0.67 & 1.80 & $-$1.62 &   0.26$\pm$0.03 &   0.49$\pm$0.04 & 0.44$\pm$0.04 & 0.42$\pm$0.02 &  8.5$\pm$0.2 & 0.57$\pm$0.02 \\
200031 & 4300 & 0.77 & 1.55 & $-$1.85 &   0.31$\pm$0.03 &   0.18$\pm$0.04 & 0.34$\pm$0.04 & 0.20$\pm$0.02 &  9.4$\pm$0.2 & 0.51$\pm$0.02 \\
200051 & 4260 & 0.90 & 1.60 & $-$1.63 &   0.73$\pm$0.03 &$-$0.05$\pm$0.04 & 0.44$\pm$0.04 & 0.40$\pm$0.02 &  6.8$\pm$0.2 & 0.58$\pm$0.02 \\
200061 & 4430 & 1.05 & 1.70 & $-$1.78 &$-$0.17$\pm$0.03 &   0.34$\pm$0.04 & 0.54$\pm$0.04 & 0.28$\pm$0.02 & 12.1$\pm$0.6 & 0.61$\pm$0.03 \\
200068 & 4500 & 1.30 & 1.52 & $-$1.84 &$-$0.01$\pm$0.03 &   0.45$\pm$0.04 & 0.23$\pm$0.04 & 0.22$\pm$0.02 &  9.4$\pm$0.3 & 0.56$\pm$0.02 \\
\multicolumn{10}{c}{{\bf M\,30}\tablefootmark{3}} \\
5783   & 4463 & 1.17 & 2.32 & $-$2.33 & 0.57$\pm$0.02 &   0.16$\pm$0.12 &  $...$         & $...$ &  8.7$\pm$0.7 & 0.57$\pm$0.03 \\
10849  & 4365 & 0.96 & 2.34 & $-$2.40 & 0.65$\pm$0.04 &$-$0.29$\pm$0.20 & $...$       & $...$ & 11.9$\pm$0.9 & 0.86$\pm$0.04 \\
11294  & 4258 & 0.41 & 2.14 & $-$2.37 & 0.07$\pm$0.07 &   0.28$\pm$0.08 & 0.51$\pm$0.20 & $...$ & 12.2$\pm$0.4 & 0.53$\pm$0.02 \\
\hline
 \end{tabular}
\par}
\tablefoot{
\tablefoottext{1}{IDs, atmospheric parameters and abundances from \citet{marino+2008}. The adopted Solar values are as follows: $A(\rm{Fe}) = 7.48$, $A(\rm{Na}) = 6.32$, $A(\rm{O}) = 8.83$, $A(\rm{Mg}) = 7.55$, $A(\rm{Ca}) = 6.39$.}\\
\tablefoottext{2}{IDs, atmospheric parameters and abundances from \citet{marino+2009}. The adopted Solar values are as follows: $A(\rm{Fe}) = 7.48$, $A(\rm{Na}) = 6.31$, $A(\rm{O}) = 8.83$, $A(\rm{Mg}) = 7.54$, $A(\rm{Ca}) = 6.39$.}\\
\tablefoottext{3}{IDs, atmospheric parameters and abundances from \citet{carretta+2009c,carretta+2009b}. The adopted Solar values are as in \citet{gratton+2003}: $A(\rm{Fe}) = 7.54$, $A(\rm{Na}) = 6.21$, $A(\rm{O}) = 8.79$, $A(\rm{Mg}) = 7.43$, $A(\rm{Ca}) = 6.27$.}\\
\tablefoottext{4}{The macroturbulence velocity (v$_{mac}$) is a free parameter in the fitting procedure.}\\
\tablefoottext{5}{The adopted Solar values are $A(\rm{Fe}) = 7.50$, $A(\rm{S}) = 7.16$ \citep{caffau+2011}.}\\
}
\end{center}
\end{table*}

Our spectra were taken in service mode under Programme ID 091.B-0171(A) with the VLT cryogenic high-resolution infrared echelle spectrograph \citep[CRIRES][]{kaeufl+2004} mounted at the Nasmyth focus A of the UT1 telescope (Antu).
We used the $0.4\arcsec$~slit, which provides a resolving power $R\sim50000$ and a dispersion scale on the four parallel, $512$\,px CCD chips of $0.05$\,\AA\,px$^{-1}$. The S I multiplet 3 was centred on chip 3.
We used relatively short exposure times (between $30$ and $300$\,s) with a nodding cycle between each exposure. The total integration time was chosen as to achieve a SNR~$\sim 200$ per px in the region of the sulphur lines. In the nodding technique, each star is moved along the slit and exposed at two different positions. This allows for proper sky subtraction and removal of systematic effects, such as amplifier glow.
The final, combined spectra are the mean of all short exposures.
The science targets were also used as guide stars for the adaptive optics system.
The observing log is presented in Table \ref{tab:obs_log}.

\begin{table}
\begin{center}
\caption{Observing log.}\label{tab:obs_log}
{\footnotesize
 \begin{tabular}{ccccc}
\hline
 ID & V & Date & Exp. time &  SNR  \\
    & [mag] &      & [s]       &   [px$^{-1}$]    \\
\hline
\multicolumn{5}{c}{M\,4} \\
19925 & 11.04 & Jun. 23/24 2013 & $8\times60$  & 220 \\
21191 & 11.70 & Sep. 13/14 2013 & $8\times120$ & 230 \\
27448 & 11.73 & Sep. 13/14 2013 & $8\times120$ & 230 \\
28103 & 10.71 & Apr.   5/6 2013 & $6\times60$  & 240 \\
34006 & 11.87 & Sep.   6/7 2013 & $8\times120$ & 160 \\
36215 & 11.80 & Sep.   6/7 2013 & $8\times120$ & 140 \\
\multicolumn{5}{c}{M\,22} \\
200005 & 10.92 & Jun. 19/20 2013 & $8\times60$   & 230 \\
200025 & 11.52 & Jun. 19/20 2013 & $12\times60$  & 190 \\
200031 & 11.64 & Jun. 19/20 2013 & $8\times120$  & 180 \\
200051 & 12.04 & Jun. 20/21 2013 & $10\times120$ & 140 \\
200061 & 12.21 & Aug.  9/10 2013 & $14\times180$ & 100 \\
200068 & 12.30 & Aug. 10/11 2013 & $16\times120$ & 170\\
\multicolumn{5}{c}{M\,30} \\
5783\tablefootmark{1}  & 12.71 & Jul. 27/28 2013 & $8\times300$  & 140 \\
5783\tablefootmark{1}  & 12.71 & Aug. 22/23 2013 & $8\times300$  & 140 \\
10849  & 12.55 & Jul. 27/28 2013 & $12\times300$ & 100 \\
11294  & 12.09 & Aug. 22/23 2013 & $6\times300$  & 220 \\
\hline
 \end{tabular}
\par}
\tablefoot{
\tablefoottext{1}{This star was observed in two nights. The stated SNR is for the combined spectrum of all $16\times300$\,s exposures.}
}
\end{center}
\end{table}

For the data reduction, we used the standard CRIRES pipeline (version 2.3.1) provided by ESO.
The pipeline allows dark current subtraction, bad pixel correction, sky subtraction, flat-field correction, non-linearity correction, wavelength calibration from a Th-Ar lamp, and an extraction and combination of the final 1D spectrum from the science images.
We noticed that the wavelength calibration was suboptimal due to the very few Th-Ar lamp lines with sufficient brightness available in the observed region, so we additionally refined the wavelength solution using the available stellar lines in the spectra.
Finally, the spectra were continuum normalised by fitting the continuum regions with a high order polynomial.

\section{Results}

\subsection{Spectral synthesis}

\begin{figure}
   \centering
   \includegraphics[width=\hsize]{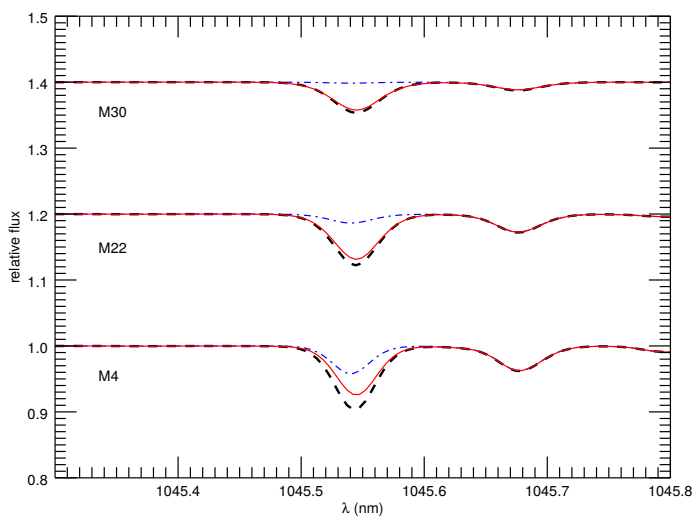}
      \caption{Synthetic spectra illustrating the relative contribution of the Fe I line (blue dash-dotted line) at $1045.5$\,nm to the S I line at the same wavelength (red line) with respect to the combined spectral feature (thick dashed black line) at the metallicities of the three GCs assuming Solar abundance mix. 
              }
         \label{fig:blend}
   \end{figure}

We derived the S abundance for each of the observed stars through a full spectral synthesis using the LTE code {\it ``Spectroscopy Made Easy''} \citep[SME][]{valenti+piskunov96}.
We derived our line list from the VALD database \citep{kupka+1999,kupka+2000}. The transition parameters of the lines in multiplet 3 and the Fe I line at $1047.0$\,nm are presented in Table \ref{tab:lines}.
There is no severe line blending in the wavelength region of multiplet 3. Only the S I line at $1045.3$\,nm is blended with a Fe I line but the latter has a marginal contribution at [Fe/H]~$\sim-1$\,dex and practically disappears at lower metallicities. The relative contribution of this Fe I line lying on top of the S I feature is shown in Figure \ref{fig:blend} for the metallicities of the three GCs.
The atomic data for this blending iron line is also given in Table \ref{tab:lines}.
All other transitions on top of the S triplet are of a negligible strength but we note that we use the complete line-list of the VALD database, including molecular bands in the respective wavelength interval in our synthesis.
The region is free of telluric absorption lines.
We interpolated the new grid of Kurucz\footnote{\url{http://kurucz.harvard.edu/grids.html}} plane-parallel, one-dimensional models without convective overshoot. These include the $\alpha$-enhanced opacity distribution functions \citep[AODFNEW;][]{castelli+kurucz2003}\footnote{\url{http://wwwuser.oats.inaf.it/castelli/}}. The models fully cover the parameter space of all stars in our sample with steps $\Delta T_{eff} = 250$\,K, $\Delta\log g = 0.5$\,dex, and $\Delta\rm{[Fe/H]} = 0.5$\,dex. Kurucz models have also been used for the derivation of the stellar parameters (Table~\ref{tab:targets}).
We further discuss the impact of different atmospheric models to the S abundances in Section 3.2.

\begin{table*}
\begin{center}
\caption{Line parameters from VALD.}\label{tab:lines}
{\small
 \begin{tabular}{cccccccc}
\hline
 Element & $\lambda$ & $\chi$ & $\log gf$ &  \multicolumn{3}{c}{Damping parameters} & Land\'e factor  \\
         & [nm]     & [eV]   &           &   Radiative     & Stark      & Waals    &               \\
\hline
S I   &    1045.5449\tablefootmark{1}  & 6.8600\tablefootmark{1} & $+$0.250\tablefootmark{1} & 8.950\tablefootmark{2} & $-$5.370\tablefootmark{2} & $-$7.333\tablefootmark{3} & 1.250\tablefootmark{2} \\
Fe I  &    1045.5404\tablefootmark{4}  & 5.3930\tablefootmark{4} & $-$0.905\tablefootmark{4} & 8.480\tablefootmark{4} & $-$3.960\tablefootmark{4} & $-$7.520\tablefootmark{4} & 1.130\tablefootmark{4} \\
S I   &    1045.6757\tablefootmark{1}  & 6.8600\tablefootmark{1} & $-$0.447\tablefootmark{1} & 8.950\tablefootmark{2} & $-$5.370\tablefootmark{2} & $-$7.333\tablefootmark{3} & 2.000\tablefootmark{2} \\
S I   &    1045.9406\tablefootmark{1}  & 6.8600\tablefootmark{1} & $+$0.030\tablefootmark{1} & 8.950\tablefootmark{2} & $-$5.370\tablefootmark{2} & $-$7.333\tablefootmark{3} & 1.750\tablefootmark{2} \\
Fe I  &    1046.9653\tablefootmark{4}  & 3.8840\tablefootmark{4} & $-$1.187\tablefootmark{4} & 6.830\tablefootmark{4} & $-$6.150\tablefootmark{4} & $-$7.820\tablefootmark{4} & 1.170\tablefootmark{4} \\

\hline
 \end{tabular}
\par}
\tablefoot{Line data references:
\tablefoottext{1}{\citet{ZCBS}}
\tablefoottext{2}{\citet{K04}}
\tablefoottext{3}{\citet{BPM}}
\tablefoottext{4}{\citet{K07}}
}
\end{center}
\end{table*}

\begin{figure}
   \centering
   \includegraphics[width=\hsize]{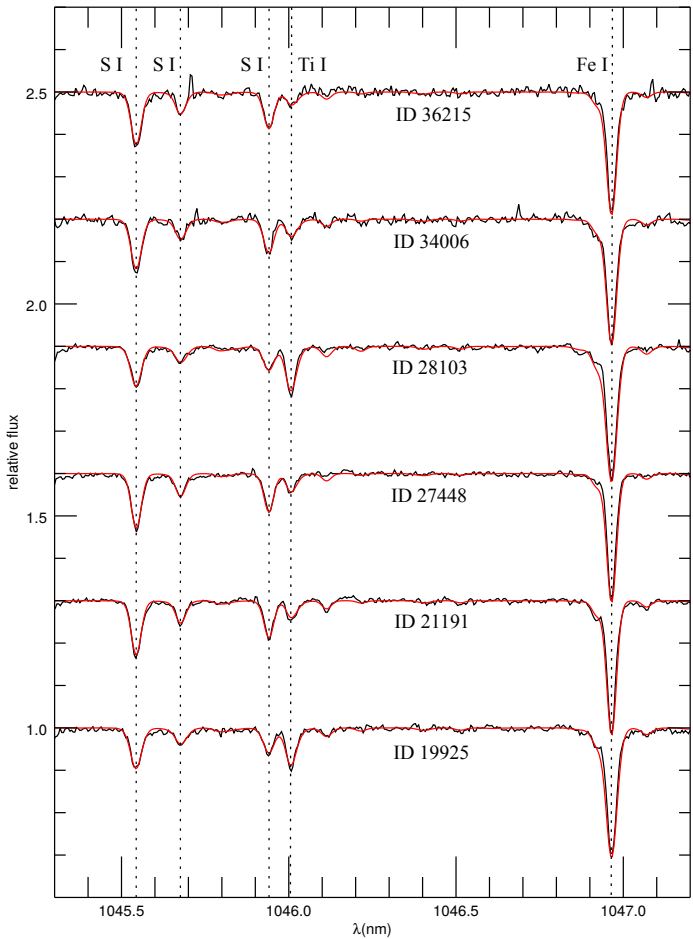}
      \caption{Observed (black) and synthetic (red) spectra for the stars in M\,4. The most prominent spectral lines are marked with vertical dotted lines. The spectra are shifted vertically by an arbitrary value.
              }
         \label{fig:m4}
   \end{figure}

\begin{figure}
   \centering
   \includegraphics[width=\hsize]{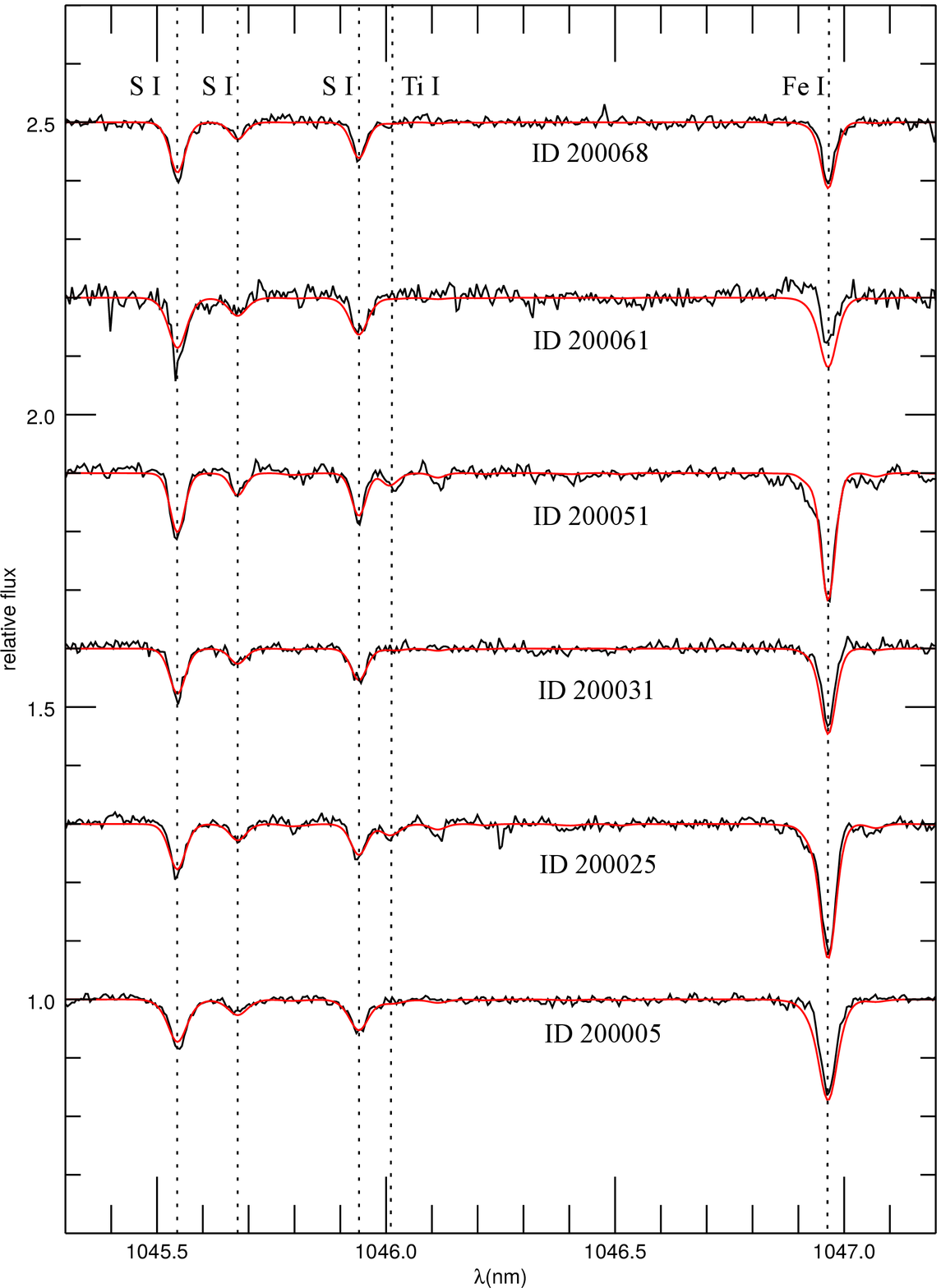}
      \caption{Same as Figure \ref{fig:m4} but for the stars in M\,22.
              }
         \label{fig:m22}
   \end{figure}

\begin{figure}
   \centering
   \includegraphics[width=\hsize]{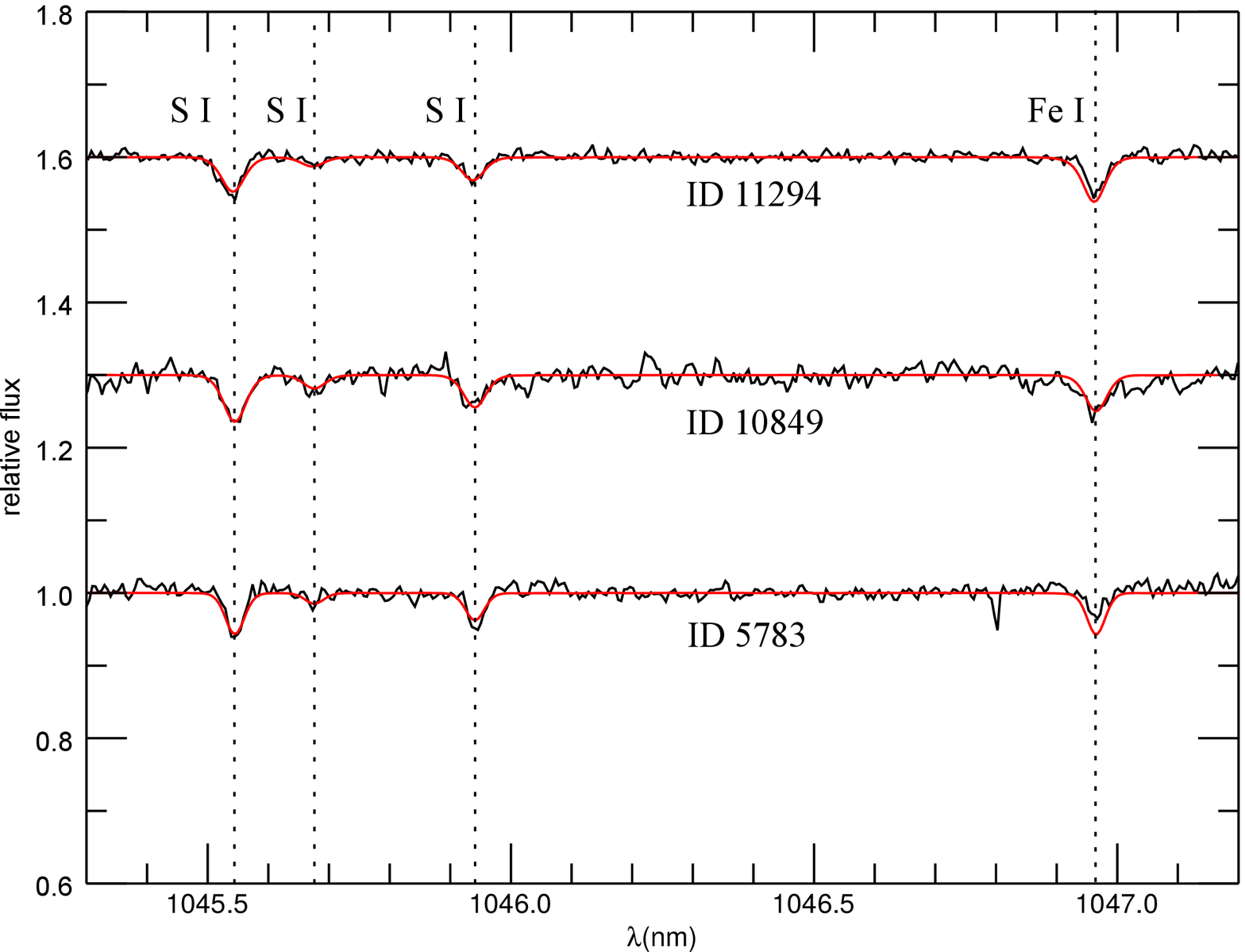}
      \caption{Same as Figure \ref{fig:m4} but for the stars in M\,30.
              }
         \label{fig:m30}
   \end{figure}

SME performs a $\chi^2$ minimisation algorithm iterating a defined set of free parameters.
In our case, we used the [S/Fe] ratio and the macroturbulence velocity of the stars as free parameters to simultaneously fit the three S lines from multiplet 3 and the only prominent Fe line in the region (see Figures \ref{fig:m4}, \ref{fig:m22}, \ref{fig:m30}).
The stellar parameters, as presented in Table \ref{tab:targets}, were kept constant.
The goodness of fit of the Fe I line shows the suitability of the input atmospheric parameters, since it is only affected by changes in the broadening parameter -- the macroturbulence velocity in this case. 
   
The estimated [S/Fe] ratio in LTE and the best-fit values for the macroturbulence velocity for each star together with the statistical uncertainties are presented in the last two columns of Table~\ref{tab:targets}.
We used the value of $A(\rm{S}) = 7.16$\,dex and $A(\rm{Fe}) = 7.50$\,dex as in \citet{caffau+2011} for the Solar abundance of sulphur and iron, respectively.

We find very homogeneous [S/Fe] ratios for all $6$ stars in M\,4 with a mean $\mathrm{[S/Fe]_{LTE}} = 0.58\pm0.01$\,dex with an intrinsic spread $\sigma_0 = 0.04$\,dex.
The intrinsic spread is calculated as $\sigma_0^2~=~\sigma_{\mathrm{[S/Fe]}}^2-<\epsilon_{rand}>^2$, where $\sigma_{\mathrm{[S/Fe]}}$ is the standard deviation of the [S/Fe] ratio from the mean of all stars and $<\epsilon_{rand}>$ is the average random error of the individual measurements.
The random errors ($\epsilon_{rand}$) are the formal uncertainties based on the goodness of the $\chi^2$ fit.
Five of the observed stars in M\,22, regardless of their metallicity, have consistent abundances with a mean $\mathrm{[S/Fe]_{LTE}} = 0.57\pm0.01$\,dex with an intrinsic spread $\sigma_0 = 0.03$\,dex. There is only one star (ID 200005) that shows somewhat higher $\mathrm{[S/Fe]_{LTE}} = 0.78\pm0.01$\,dex.
The situation is similar with M\,30. Two stars show consistent [S/Fe] ratios with a mean $\mathrm{[S/Fe]_{LTE}} = 0.55\pm0.02$\,dex and the third one (ID 10849) has an abnormally high $\mathrm{[S/Fe]_{LTE}} = 0.86\pm0.04$\,dex.
Note, however, that the derived [S/Fe] ratios are strongly dependent on the effective temperatures and moderately on the gravities, and metallicities of the stars (see Section 3.2). Thus, reasonable changes in the effective temperatures of the stars with abnormal S abundances can bring them in line with the other stars in the sample, e.g. an increase of $\mathrm{T_{eff}}$ by $100$\,K for star 200005 in M\,22 and by $200$\,K for star 10849 in M\,30.
Star 200005 also has a somewhat lower [Fe/H] estimate $\sim0.1$\,dex lower than the mean of the metal-poor population of M\,22 \citep{marino+2009}. As it is indicated in Table \ref{tab:systematic}, increasing the metallicity of the star by $0.1$\,dex will bring the [S/Fe] ratio down by $0.1$\,dex, already in a reasonable agreement with the rest of the stars in M\,22.

The reference Fe I line in our spectra for star 200005 is also not well fit by the synthetic spectrum, with a $\chi^2$ estimation about $2.5$~times higher than the average for the other stars and cannot be improved by only changing the macroturbulence velocity. This means that the model spectrum does not reproduce correctly the equivalent width of the line, which is another indication that the stellar parameters for this star might be off. We are able to obtain very good fits to the iron line by modifying the effective temperature, gravity, and metallicity of the star but due to the degeneracy between these parameters, it is difficult to put any firm conclusions on its sulphur abundance.
The situation with star 10849 in M\,30 (it also has the lowest SNR spectrum in our sample) is more critical.
As our comparison of the stellar parameter determination using the methodology of \citet{carretta+2009b} and \citet{marino+2008} for the common stars in M\,4 showed, there are systematic differences but the standard deviations in $\rm{T_{eff}}$ and $\log  g$ appear to be small.
We do not have solid reasons to believe that the relative estimate of the effective temperature of a star in a homogeneous study could be off by as much as $200$\,K and it does not present odd abundance of any other element \citep{carretta+2009b}.

\subsection{Systematic uncertainties}

   \begin{figure}
   \centering
   \includegraphics[width=\hsize]{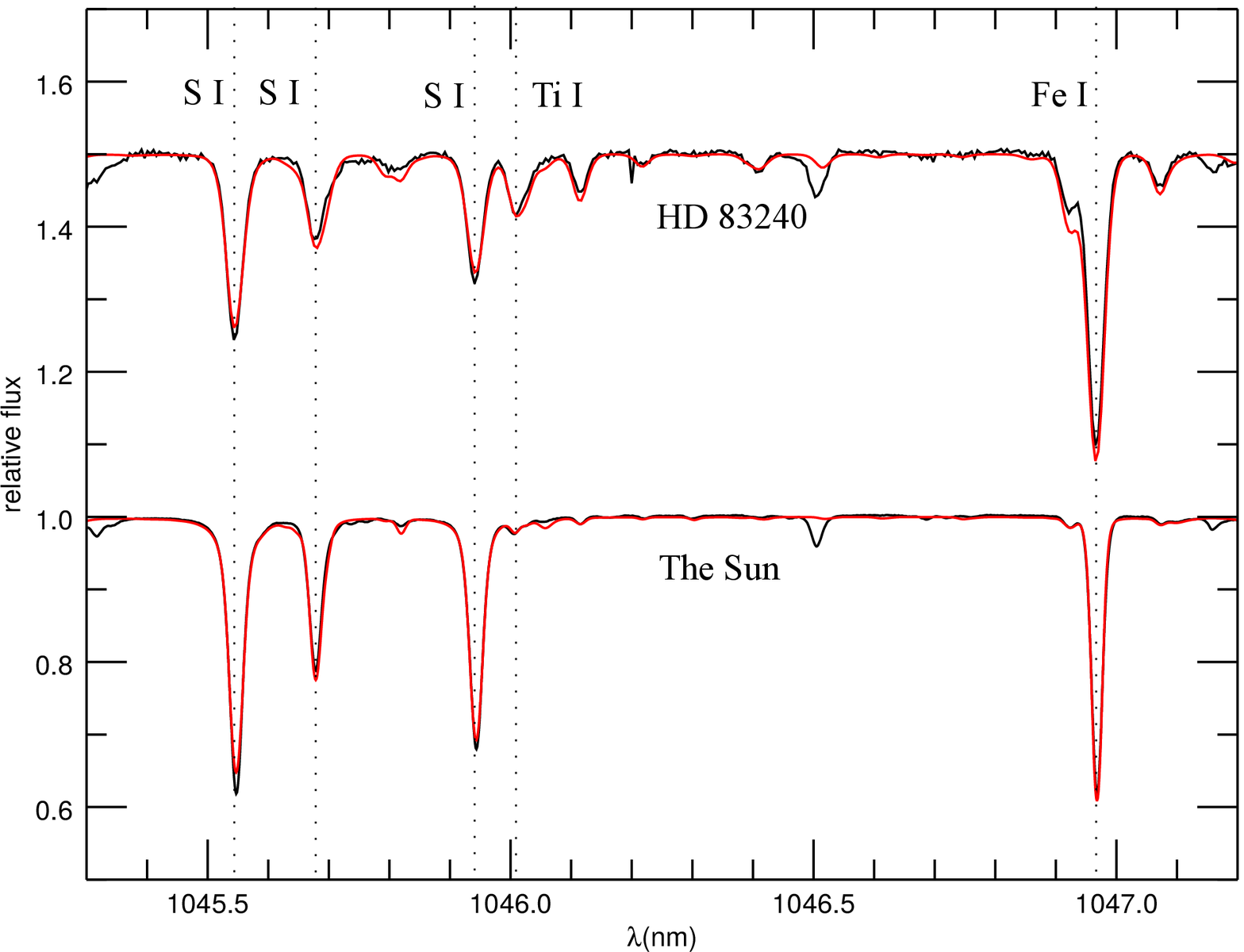}
      \caption{The S multiplet 3 region in the Sun and the benchmark RGB star HD\,83240 fitted using the same technique as for the GC stars. In the Sun, we estimate $A(S)_{\mathrm{LTE}}=7.26$\,dex opposed to the adopted Solar abundance of $A(S)=7.16$\,dex. For HD\,83240 we find $A(S)_{\mathrm{LTE}}=7.13$\,dex. Symbols are as in Figure \ref{fig:m4}.
              }
         \label{fig:sun}
   \end{figure}

In order to estimate the systematic errors of the [S/Fe] ratio caused by uncertainties in the atmospheric parameters of the stars, we chose one star from each GC and varied its $\mathrm{T_{eff}}$ by $\pm100$\,K, $\log g$ by $\pm0.1$\,dex, $v_{mic}$ by $\pm0.1$\,\kms, and [Fe/H] by $\pm0.05$\,dex.
These values are very similar to the uncertainties of the atmospheric parameters given in the literature. 
The results are presented in Table \ref{tab:systematic}. In the last column, we list the total systematic error, which is calculated by combining in quadrature the uncertainties caused by different parameters. We note, however, that this is an upper limit of the systematic uncertainties due to covariances of the different parameters \citep[see][]{mcwilliam+95}.

\begin{table*}
\begin{center}
\caption{Systematic uncertainties of [S/Fe].}\label{tab:systematic}
{\small
 \begin{tabular}{ccccccccccc}
\hline
GC & Star ID & \multicolumn{2}{c}{$\Delta T_{eff}$} & \multicolumn{2}{c}{$\Delta \log g$} & \multicolumn{2}{c}{$\Delta v_{mic}$} & \multicolumn{2}{c}{$\Delta$[Fe/H]} & total \\
 &  & $+100$\,K & $-100$\,K & $+0.1$\,dex & $-0.1$\,dex & $+0.1$\,\kms & $-0.1$\,\kms & $+0.05$\,dex & $-0.05$\,dex \\      
\hline
M\,4  & 27448  & $-0.18$ & $+0.20$ & $+0.04$ & $-0.05$ & $-0.01$ & $ 0.00$ & $-0.03$ & $+0.03$ & $\pm0.20$ \\
M\,22 & 200051 & $-0.18$ & $+0.18$ & $+0.05$ & $-0.06$ & $ 0.00$ & $ 0.01$ & $-0.05$ & $+0.03$ & $\pm0.19$ \\
M\,30 & 11294  & $-0.14$ & $+0.16$ & $+0.05$ & $-0.04$ & $ 0.00$ & $ 0.01$ & $-0.02$ & $+0.03$ & $\pm0.16$ \\
\hline
 \end{tabular}
\par}
\end{center}
\end{table*}

The choice of the atmosphere structure also has a marginal impact on the derived abundances. For instance, using a Solar-scaled Kurucz atmosphere model (ODFNEW) would lead to $0.12$\,dex lower on average S abundances for the stars in M\,4. The differences are less pronounced in the more metal-poor GCs where the Solar-scaled models would render S abundances $0.07$\,dex and $0.04$\,dex lower for M\,22 and M\,30, respectively.
We also tested the Marcs 2012 spherically symmetric models, which are $\alpha$-enhanced for metallicities below $-1$\,dex \citep{gustafsson+2008}. These models lead to S abundances $0.07$\,dex higher on average than the ones stated in Table \ref{tab:targets} for all three GCs.
All model grids are equally densely populated in the parameter space of interest.

\subsection{NLTE and 3D effects}

The S lines at $1045$\,nm are affected by relatively large NLTE effects. By extrapolating the NLTE results from the grid of \citet{takeda+2005} to the stellar parameters of our stars, we found $\Delta_{NLTE} = -0.20$\,dex, $\Delta_{NLTE} = -0.15$\,dex, and $\Delta_{NLTE} = -0.10$\,dex for the GCs M\,4, M\,22, and M\,30, respectively.
The corrections become smaller for lower metallicities.
This produces a weak trend of slightly increasing $\rm{[S/Fe]_{NLTE}}$ ratios with decreasing metallicity with respect to the very homogeneous mean $\rm{[S/Fe]_{LTE}}$ values for the three GCs stated in Section 3.1. We find $\rm{[S/Fe]_{NLTE}} = 0.38$\,dex, $0.42$\,dex, and $0.45$\,dex for the GC M\,4, M\,22, and M\,30, respectively -- more in-line with the nominal $\rm{[\alpha/Fe]} = 0.4$\,dex value for the old halo stellar population.
Considering the large systematic uncertainties, the small increase of the $\rm{[S/Fe]_{NLTE}}$ ratios with decreasing metallicity is probably insignificant.

According to calculations by \citet{jonsson+2011}, based on the 3D model atmospheres of \citet{collet+2007,collet+2009}, the 3D corrections for S abundances derived from multiplet 3 are constant and positive at all metallicities for stars with similar parameters to ours and are in the order of $\mathrm{\Delta_{3D}} = +0.20$\,dex.
\citet{caffau+2007,caffau+2010} explored the 3D effects of multiplet 3 in dwarf stars and also found positive corrections based on the CO$^5$BOLD 3D model atmospheres. Although it is inconsistent to apply their results to giant stars, there is an agreement that the 3D corrections for the lines at $1045$\,nm are always positive.
In general, the NLTE and 3D corrections for our stars roughly compensate each other and are in the order of the systematic uncertainties.
One, however, needs to be conscious when applying 3D-LTE corrections to 1D-NLTE corrected results since the abundances derived from full 3D-NLTE analysis might differ substantially (see Section 3.4).
For this reason, and to be consistent with literature [S/Fe] measurements of field and cluster stars, we consider further in the discussion only the $\rm{[S/Fe]_{NLTE}}$ results stated above.

\subsection{Sulphur abundance of the Sun and HD83240}

Finally, the [S/Fe] ratio is relative to the Solar abundance of sulphur and therefore it is vital for the proper analysis. We decided to apply the same spectral synthesis technique to the same region of the Kurucz Solar atlas \citep{kurucz+1984} and derive the LTE Solar abundance of sulphur estimated from multiplet~3 (Figure~\ref{fig:sun}).
We found $A(S)=7.26\pm0.01$\,dex, which is by $0.10$\,dex higher than the nominal value of $A(S)=7.16$\,dex \citep{caffau+2011} adopted in this work. Our result is, however, consistent with the Solar value estimated by \citet{caffau+2007} using the same lines of multiplet 3 ($A(S)=7.30$\,dex). This finding will systematically decrease our derived abundances in LTE in the three GCs by $0.1$\,dex.
Although the 1D-NLTE and 3D corrections for the Solar abundance derived from multiplet 3 are small and compensating each other -- $\Delta_{NLTE} = -0.07$\,dex \citep{takeda+2005} and $\Delta_{3D} = +0.07$\,dex \citep{caffau+2007}, it is likely that a NLTE analysis performed in a full 3D synthesis will bring the abundance down to the nominal value.
The nominal Solar value is estimated from the S triplet at $675.7$\,nm and the [S I] line at $1082$\,nm, which are virtually unaffected by NLTE effects \citep{caffau+2007,caffau+ludwig2007}.

In order to test the reliability of our results for RGB GC stars we also carried out an identical analysis for a well studied giant benchmark star -- HD\,83240. We used a CRIRES spectrum from the library of \citet{lebzelter+2012}\footnote{\url{http://www.univie.ac.at/crirespop/index.html}}. HD\,83240 is classified as a K1\,III star and has a $\rm{T_{eff}} = 4680$\,K, $\rm{\log g} = 2.45$\,dex, $\rm{v_{mic}} = 1.3$\,\kms~and $\rm{[Fe/H]}=-0.02$\,dex. Although, no sulphur abundance was explicitly derived for this star, \citet{mishenina+2006,mishenina+2007} measured the abundances of various light, Fe-peak, and n-capture elements and found that the star has a Solar abundance pattern with a slight under-abundance of carbon and over-abundance of nitrogen, consistent with a post first-dredge-up evolutionary stage.
Thus, we assume that HD\,83240 also has Solar abundance of sulphur.
We measured $\rm{[S/Fe]_{LTE}} = -0.01$\,dex (Figure \ref{fig:sun}).
The NLTE correction for this star is expected to be in the order of $-0.15$\,dex and the 3D correction is positive and in the same order of magnitude.

The Fe I line that we use to adjust the broadening in our spectra is well reproduced by our synthesis in both the Sun and HD\,83240. Thus we confirm the suitability of the atomic data for this line.

\section{Implications on the Galactic chemical evolution of sulphur}

   \begin{figure*}
   \centering
   \includegraphics[width=17cm]{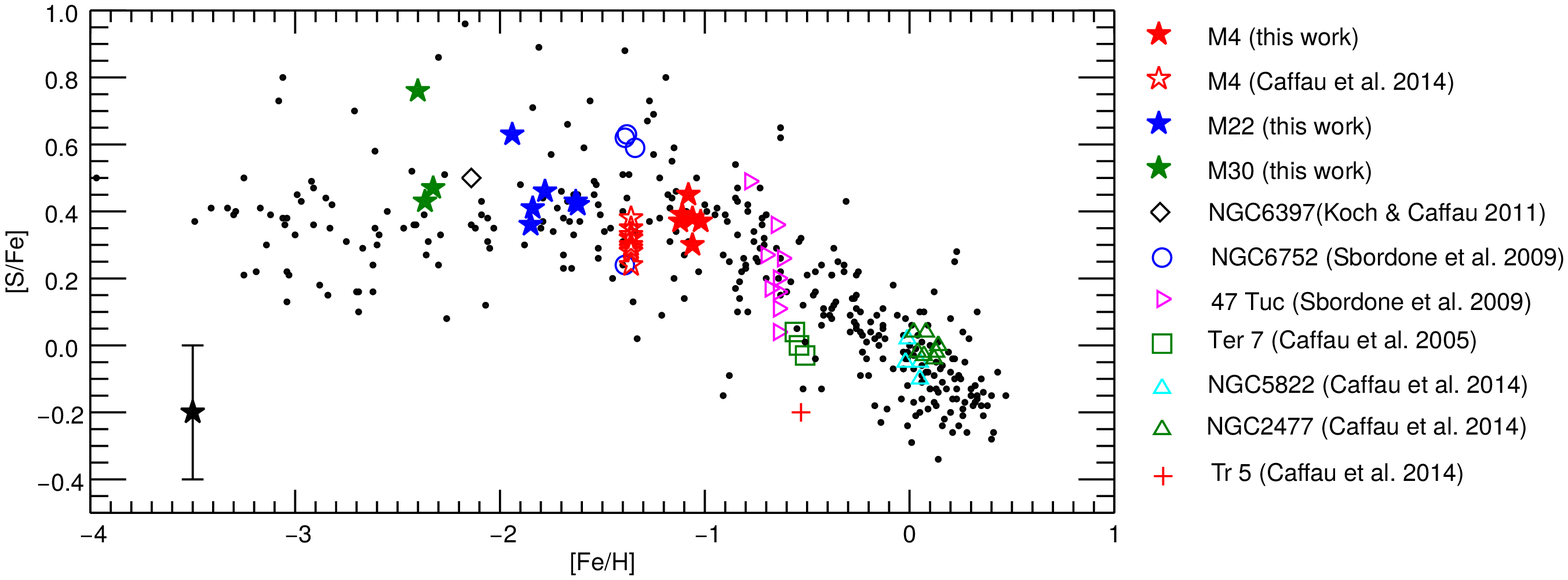}
      \caption{NLTE corrected [S/Fe] abundance ratios as a function of metallicity in Galactic disk and halos stars \citep[black dots:][]{caffau+2005,jonsson+2011,takeda+takada-hidai2011,spite+2011} and cluster stars (colourful symbols). All abundances are referenced to the adopted Solar value of $A(S) = 7.16$\,dex. The random errors of our GC stars are comparable to the symbol sizes. The maximum systematic error due to uncertainties of the stellar parameters is shown in the figure.
              }
         \label{fig:s_nlte}
   \end{figure*}
   
   \begin{figure*}
   \centering
   \includegraphics[width=\hsize]{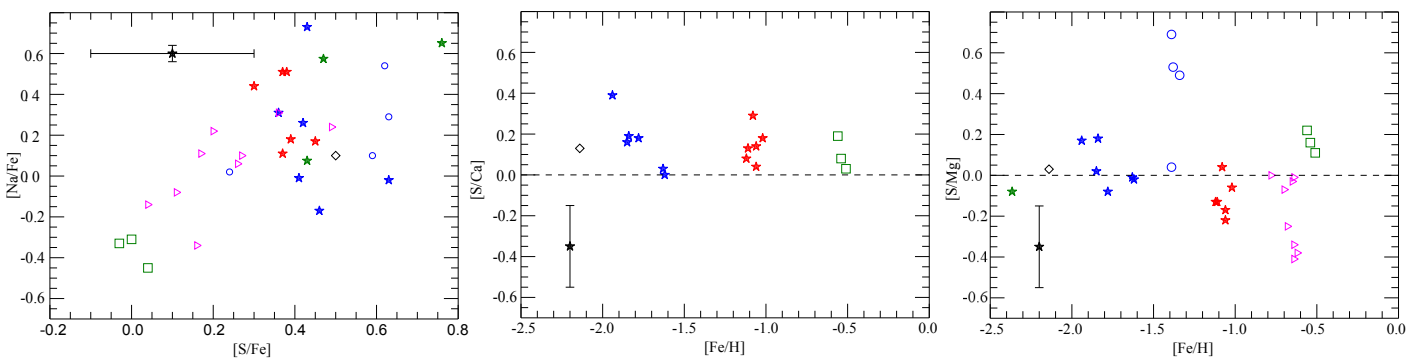}
      \caption{A comparison of the S abundances with the abundances of other light and $\alpha$ elements in GCs: [S/Fe] vs. [Na/Fe] ({\it left panel}); [Fe/H] vs. [S/Ca] ({\it middle panel}); [Fe/H] vs. [S/Mg] ({\it right panel}). Different symbols indicate stars from different clusters as indicated in the legend of Figure \ref{fig:s_nlte}. The maximum systematic errors are also indicated.
              }
         \label{fig:s_alpha}
   \end{figure*}

We should be careful when directly comparing the stellar [S/Fe] abundance ratios from different studies because they are based on different methods, different S lines, and different assumptions in their final results. It has been shown that even after proper NLTE treatment, the various S features could result in offsets in the derived abundances \citep[see][]{jonsson+2011}.
Having said that, in Figure \ref{fig:s_nlte} we have compiled a large collection of known [S/Fe] ratios (references are provided in the figure legend and caption) for Galactic and cluster stars in order to roughly track the evolution of sulphur. All measurements have been shifted to the Solar value of sulphur ($A(S) = 7.16$\,dex) adopted in this work and NLTE corrections are applied where applicable.
Most of the NLTE corrections come from the calculations of \citet{takeda+2005}, except for the extremely metal poor stars of \citet{spite+2011} who compute their own NLTE corrections. We note that, although the NLTE results by both teams are computed for different types of stars, there are some differences in the overlapping regime.
We did not apply 3D corrections in the results plotted in Figure \ref{fig:s_nlte} in order to be consistent with most of the literature discussions.

There are S abundances derived within the entire span of metallicities covered by stars in Galactic clusters, although the sample is still limited and further studies are desirable.
In all analysed clusters, sulphur seems to behave as a typical $\alpha$ element, meaning that the stars in metal poor Galactic GCs ([Fe/H]~$<-1.0$\,dex) have enhanced [S/Fe] abundance ratios and open clusters of solar metallicity have solar [S/Fe] ratios, consistent with the results for field stars and chemical evolutionary models. The only exceptions are Tr\,5 discussed in \citet{caffau+2014} and Ter\,7 \citep{caffau+2005b} with anomalously low [S/Fe] ratios. The latter is associated with the Sagittarius dwarf spheroidal galaxy and its low [S/Fe] ratio is consistent with the lower star formation efficiency of its parent galaxy.
While the spread of sulphur on the metal poor plateau as traced by cluster stars is smaller on average than the spread traced by Galactic field stars we found one star in M\,22 and one in M\,30 that are compatible with extreme S abundances.
   
   
Messier\,4 is the only GC analysed by two teams independently \citep[][and this work]{caffau+2014} and a more careful comparison between the results is possible. \citep{caffau+2014} derive S abundances for 10 stars using line profile fitting of multiplet 1 at $920$\,nm taking also into account the telluric absorption lines.
Although, both studies seem to agree on consistent [S/Fe] abundance ratios for this GC (see Figure \ref{fig:s_nlte}), \citet{caffau+2014} assume lower metallicity for their stars, adopting [Fe/H]~$=-1.36$\,dex for M\,4 \citep{monaco+2012}, while the mean [Fe/H] for our stars is $=-1.08$\,dex \citep{marino+2008}. 
This discrepancy results into much lower value of $A(S) = 6.11\pm0.04$\,dex estimated by \citet{caffau+2014}, compared to our estimate of $A(S) = 6.46\pm0.02$\,dex in NLTE. The NLTE corrections for both multiplets are similar and in the order of $0.2$\,dex \citep{takeda+2005}.
Both studies agree, however, that there is not any significant spread in [S/Fe] amongst the RGB stars in M\,4.

Finally, we explore the behaviour of sulphur with respect to other light and $\alpha$ elements. One of the goals of this study is to probe whether star-to-star variations within an individual GC are possible as is the case for elements like C, N, O, Na, Al, and to some extent Mg and Si, which are all correlated between each other \citep[see the review by][and references therein]{gratton+2012}.
In Figure \ref{fig:s_alpha} (left panel), we plot existing measurements of [S/Fe] vs. [Na/Fe] for various GCs.
In M\,4, M\,22, and M\,30, we have deliberately targeted stars with different Na abundances and Figure \ref{fig:s_alpha} clearly shows that there is not a S-Na correlation between stars from these GCs.
While the high S star in M\,30 is also Na-rich, the high S star in M\,22 has a low Na abundance.
So far, with the only tentative exception of 47\,Tuc, already discussed in \citet{sbordone+2009}, it seems that even if present, star-to-star variations of the sulphuric abundance in GCs are not caused by the same self-enrichment processes responsible for the formation of p-capture element abundance variations.
We note, however, that we still lack an extensive study that includes a large number of stars in a single GC.
   
In the middle panel of Figure \ref{fig:s_alpha} we show the behaviour of the [S/Ca] ratio in GCs -- two $\alpha$ elements carved through the same production channels. The mean ratio is $\sim0.1$\,dex and there is no trend with metallicity, which suggests that the two elements do evolve in a lock-step, as expected.
There is more scatter in the [S/Mg] ratio, although the mean value is $0$\,dex (Figure \ref{fig:s_alpha}, right panel).
Both elements have slightly different production chains, so the larger scatter is not surprising.
In the pre-supernova evolution, Mg is produced in the C-convective shell burning, while S - in the O-convective shell burning. During the explosive nucleosynthesis, Mg is produced in C and Ne burning, while S - in O burning \citep{limongi+chieffi2003}. Magnesium could also participate in p-capture reactions through the MgAl hydrogen burning cycle at high temperatures and hence star-to-star variations of this element are prominent in some GCs.
While in a majority of the observed GC stars covering a large range of metallicities, the behaviour of sulphur with respect to other $\alpha$ elements favours a typical $\alpha$-capture origin of this element, the detection of two very S-rich stars (if confirmed real) may rise the question of additional production channels.
   
\section{Conclusions}

Due to the lack of suitable lines to measure sulphur in the visible band at low metallicities, the S abundance measurements in GC stars are quite scarce to date. The present work helps to substantially increase the sample of cluster stars with determined S abundances.
We measured the [S/Fe] abundance ratio from multiplet 3 at $1045$\,nm in 15 RGB stars from three GCs (M\,4, M\,22, and M\,30) observed with the high-resolution, infrared spectrograph CRIRES, mounted at the VLT.
The multiplet 3 is suited for S abundance derivation since the three lines are relatively strong even at low metallicity and are free of significant stellar blends and telluric contamination.
The selected GCs cover a large range of metallicities ($-2.3<$~[Fe/H]~$<-1.0$\,dex) and the individual RGB targets were specially selected to have different Na \& O abundances or even different Fe abundances in the case of the massive GC M\,22. This ensured that we have stars belonging to different cluster populations and allowed us to search for possible star-to-star variations of the S abundance in GCs.

We find that the S abundances derived from multiplet 3 are very sensitive to changes in the effective temperature of the observed stars and the uncertainty of this parameter dominates the uncertainty of the measured abundances.
The lines of multiplet 3 are also prone to large departures from LTE and are sensitive to horizontal fluctuations of the stellar atmosphere, inferring large 3D model corrections.
The NLTE and 3D corrections are similar in magnitude and roughly compensate each other, however the lack of full NLTE analysis performed within the framework of a 3D atmospheric model introduces additional systematic uncertainty.
Nevertheless, we perceive very small star-to-star scatter in the majority of the observed stars in the three GCs consistent with a narrow plateau ($\rm{[S/Fe]_{LTE}} = 0.57\pm0.01$\,dex, $\sigma_0=0.03$\,dex; $\rm{[S/Fe]_{NLTE}} = 0.40$\,dex, $\sigma_0=0.05$\,dex), as expected for the behaviour of a typical $\alpha$ element at low metallicity in the Milky Way.
Sulphur appears to vary in a lock step with Ca, a well studied $\alpha$ element most likely produced through the same nuclear channels. The mean [S/Mg] ratio is Solar but with a larger scatter due to the more complex production path of Mg, which could also participate in the self-enrichment processes that take place in GCs.

One star in the GC M\,22 and one star in the GC M\,30 are found to be exceptionally S rich, being consistent with the high S measurements in some field stars \citep[e.g.][]{caffau+2005}.  We consider the possibility that these are erroneous measurements due to some deviation from the true atmospheric parameters of the two stars but we cannot undoubtedly exclude the possibility of observing genuine S-rich stars in GCs.
The S-rich stars do not show clear correlations with p-capture elements like Na and O, so we conclude that S is not produced during hot hydrogen burning reactions responsible for the formation of multiple population in GCs.
Discovering stars with confirmed extreme [S/Fe] ratios in different stellar environments would require the invention of new production mechanisms for sulphur.

Clearly more observations are needed, in particular large homogeneous samples, to conclusively constrain the S abundances in GCs.

\begin{acknowledgements}
We thank N. Piskunov and G. Ruchti for help with the SME code and A. O. Thygesen and H.-G. Ludwig for insightful discussions.
NK and AK acknowledge the Deutsche Forschungsgemeinschaft for funding from  Emmy-Noether grant  Ko 4161/1.
EC is grateful to the FONDATION MERAC for funding her fellowship.
\end{acknowledgements}

\bibliographystyle{aa}
\bibliography{mybiblio_v5}

\begin{thebibliography}{57}
\expandafter\ifx\csname natexlab\endcsname\relax\def\natexlab#1{#1}\fi

\bibitem[{{Barklem} {et~al.}(2000){Barklem}, {Piskunov}, \& {O'Mara}}]{BPM}
{Barklem}, P.~S., {Piskunov}, N., \& {O'Mara}, B.~J. 2000, Astron. and
  Astrophys. Suppl. Ser., 142, 467, (BPM)

\bibitem[{{Bowen} {et~al.}(2005){Bowen}, {Jenkins}, {Pettini}, \&
  {Tripp}}]{bowen+2005}
{Bowen}, D.~V., {Jenkins}, E.~B., {Pettini}, M., \& {Tripp}, T.~M. 2005, \apj,
  635, 880

\bibitem[{{Caffau} {et~al.}(2005{\natexlab{a}}){Caffau}, {Bonifacio},
  {Faraggiana}, {Fran{\c c}ois}, {Gratton}, \& {Barbieri}}]{caffau+2005}
{Caffau}, E., {Bonifacio}, P., {Faraggiana}, R., {et~al.} 2005{\natexlab{a}},
  \aap, 441, 533

\bibitem[{{Caffau} {et~al.}(2005{\natexlab{b}}){Caffau}, {Bonifacio},
  {Faraggiana}, \& {Sbordone}}]{caffau+2005b}
{Caffau}, E., {Bonifacio}, P., {Faraggiana}, R., \& {Sbordone}, L.
  2005{\natexlab{b}}, \aap, 436, L9

\bibitem[{{Caffau} {et~al.}(2007){Caffau}, {Faraggiana}, {Bonifacio}, {Ludwig},
  \& {Steffen}}]{caffau+2007}
{Caffau}, E., {Faraggiana}, R., {Bonifacio}, P., {Ludwig}, H.-G., \& {Steffen},
  M. 2007, \aap, 470, 699

\bibitem[{{Caffau} \& {Ludwig}(2007)}]{caffau+ludwig2007}
{Caffau}, E. \& {Ludwig}, H.-G. 2007, \aap, 467, L11

\bibitem[{{Caffau} {et~al.}(2011){Caffau}, {Ludwig}, {Steffen}, {Freytag}, \&
  {Bonifacio}}]{caffau+2011}
{Caffau}, E., {Ludwig}, H.-G., {Steffen}, M., {Freytag}, B., \& {Bonifacio}, P.
  2011, \solphys, 268, 255

\bibitem[{{Caffau} {et~al.}(2014){Caffau}, {Monaco}, {Spite}, {Bonifacio},
  {Carraro}, {Ludwig}, {Villanova}, {Beletsky}, \& {Sbordone}}]{caffau+2014}
{Caffau}, E., {Monaco}, L., {Spite}, M., {et~al.} 2014, \aap, 568, A29

\bibitem[{{Caffau} {et~al.}(2010){Caffau}, {Sbordone}, {Ludwig}, {Bonifacio},
  \& {Spite}}]{caffau+2010}
{Caffau}, E., {Sbordone}, L., {Ludwig}, H.-G., {Bonifacio}, P., \& {Spite}, M.
  2010, Astronomische Nachrichten, 331, 725

\bibitem[{Carretta {et~al.}(2009{\natexlab{a}})Carretta, Bragaglia, Gratton, \&
  Lucatello}]{carretta+2009c}
Carretta, E., Bragaglia, A., Gratton, R., \& Lucatello, S. 2009{\natexlab{a}},
  A\&A, 505, 139

\bibitem[{Carretta {et~al.}(2009{\natexlab{b}})Carretta, Bragaglia, Gratton,
  Lucatello, Catanzaro, Leone, Bellazzini, Claudi, D'Orazi, Momany, Ortolani,
  Pancino, Piotto, Recio-Blanco, \& Sabbi}]{carretta+2009b}
Carretta, E., Bragaglia, A., Gratton, R.~G., {et~al.} 2009{\natexlab{b}}, A\&A,
  505, 117

\bibitem[{Castelli \& Kurucz(2003)}]{castelli+kurucz2003}
Castelli, F. \& Kurucz, R.~L. 2003, IAUS, 210, 20

\bibitem[{{Centuri{\'o}n} {et~al.}(2000){Centuri{\'o}n}, {Bonifacio}, {Molaro},
  \& {Vladilo}}]{centurion+2000}
{Centuri{\'o}n}, M., {Bonifacio}, P., {Molaro}, P., \& {Vladilo}, G. 2000,
  \apj, 536, 540

\bibitem[{{Collet} {et~al.}(2007){Collet}, {Asplund}, \&
  {Trampedach}}]{collet+2007}
{Collet}, R., {Asplund}, M., \& {Trampedach}, R. 2007, \aap, 469, 687

\bibitem[{{Collet} {et~al.}(2009){Collet}, {Nordlund}, {Asplund}, {Hayek}, \&
  {Trampedach}}]{collet+2009}
{Collet}, R., {Nordlund}, {\AA}., {Asplund}, M., {Hayek}, W., \& {Trampedach},
  R. 2009, \memsai, 80, 719

\bibitem[{{Da Costa} {et~al.}(2009){Da Costa}, {Held}, {Saviane}, \&
  {Gullieuszik}}]{dacosta+2009}
{Da Costa}, G.~S., {Held}, E.~V., {Saviane}, I., \& {Gullieuszik}, M. 2009,
  \apj, 705, 1481

\bibitem[{{Garnett}(1989)}]{garnett1989}
{Garnett}, D.~R. 1989, \apj, 345, 282

\bibitem[{Gratton {et~al.}(2012)Gratton, Carretta, \& Bragaglia}]{gratton+2012}
Gratton, R.~G., Carretta, E., \& Bragaglia, A. 2012, A\&ARv, 20, 50

\bibitem[{{Gratton} {et~al.}(2003){Gratton}, {Carretta}, {Claudi}, {Lucatello},
  \& {Barbieri}}]{gratton+2003}
{Gratton}, R.~G., {Carretta}, E., {Claudi}, R., {Lucatello}, S., \& {Barbieri},
  M. 2003, \aap, 404, 187

\bibitem[{{Gustafsson} {et~al.}(2008){Gustafsson}, {Edvardsson}, {Eriksson},
  {J{\o}rgensen}, {Nordlund}, \& {Plez}}]{gustafsson+2008}
{Gustafsson}, B., {Edvardsson}, B., {Eriksson}, K., {et~al.} 2008, \aap, 486,
  951

\bibitem[{{Hendricks} {et~al.}(2014){Hendricks}, {Koch}, {Lanfranchi},
  {Boeche}, {Walker}, {Johnson}, {Pe{\~n}arrubia}, \&
  {Gilmore}}]{hendricks+2014}
{Hendricks}, B., {Koch}, A., {Lanfranchi}, G.~A., {et~al.} 2014, \apj, 785, 102

\bibitem[{{J{\"o}nsson} {et~al.}(2011){J{\"o}nsson}, {Ryde}, {Nissen},
  {Collet}, {Eriksson}, {Asplund}, \& {Gustafsson}}]{jonsson+2011}
{J{\"o}nsson}, H., {Ryde}, N., {Nissen}, P.~E., {et~al.} 2011, \aap, 530, A144

\bibitem[{{Kaeufl} {et~al.}(2004){Kaeufl}, {Ballester}, {Biereichel},
  {Delabre}, {Donaldson}, {Dorn}, {Fedrigo}, {Finger}, {Fischer}, {Franza},
  {Gojak}, {Huster}, {Jung}, {Lizon}, {Mehrgan}, {Meyer}, {Moorwood}, {Pirard},
  {Paufique}, {Pozna}, {Siebenmorgen}, {Silber}, {Stegmeier}, \&
  {Wegerer}}]{kaeufl+2004}
{Kaeufl}, H.-U., {Ballester}, P., {Biereichel}, P., {et~al.} 2004, in Society
  of Photo-Optical Instrumentation Engineers (SPIE) Conference Series, Vol.
  5492, Ground-based Instrumentation for Astronomy, ed. A.~F.~M. {Moorwood} \&
  M.~{Iye}, 1218--1227

\bibitem[{{Koch} \& {Caffau}(2011)}]{koch+caffau2011}
{Koch}, A. \& {Caffau}, E. 2011, \aap, 534, A52

\bibitem[{{Kupka} {et~al.}(1999){Kupka}, {Piskunov}, {Ryabchikova}, {Stempels},
  \& {Weiss}}]{kupka+1999}
{Kupka}, F., {Piskunov}, N., {Ryabchikova}, T.~A., {Stempels}, H.~C., \&
  {Weiss}, W.~W. 1999, \aaps, 138, 119

\bibitem[{{Kupka} {et~al.}(2000){Kupka}, {Ryabchikova}, {Piskunov}, {Stempels},
  \& {Weiss}}]{kupka+2000}
{Kupka}, F.~G., {Ryabchikova}, T.~A., {Piskunov}, N.~E., {Stempels}, H.~C., \&
  {Weiss}, W.~W. 2000, Baltic Astronomy, 9, 590

\bibitem[{{Kurucz}(2004)}]{K04}
{Kurucz}, R.~L. 2004, Robert L. Kurucz on-line database of observed and
  predicted atomic transitions

\bibitem[{{Kurucz}(2007)}]{K07}
{Kurucz}, R.~L. 2007, Robert L. Kurucz on-line database of observed and
  predicted atomic transitions

\bibitem[{{Kurucz} {et~al.}(1984){Kurucz}, {Furenlid}, {Brault}, \&
  {Testerman}}]{kurucz+1984}
{Kurucz}, R.~L., {Furenlid}, I., {Brault}, J., \& {Testerman}, L. 1984, {Solar
  flux atlas from 296 to 1300 nm}

\bibitem[{{Lebzelter} {et~al.}(2012){Lebzelter}, {Seifahrt}, {Uttenthaler},
  {Ramsay}, {Hartman}, {Nieva}, {Przybilla}, {Smette}, {Wahlgren}, {Wolff},
  {Hussain}, {K{\"a}ufl}, \& {Seemann}}]{lebzelter+2012}
{Lebzelter}, T., {Seifahrt}, A., {Uttenthaler}, S., {et~al.} 2012, \aap, 539,
  A109

\bibitem[{{Limongi} \& {Chieffi}(2003)}]{limongi+chieffi2003}
{Limongi}, M. \& {Chieffi}, A. 2003, \apj, 592, 404

\bibitem[{{Marino} {et~al.}(2009){Marino}, {Milone}, {Piotto}, {Villanova},
  {Bedin}, {Bellini}, \& {Renzini}}]{marino+2009}
{Marino}, A.~F., {Milone}, A.~P., {Piotto}, G., {et~al.} 2009, \aap, 505, 1099

\bibitem[{{Marino} {et~al.}(2011){Marino}, {Sneden}, {Kraft}, {Wallerstein},
  {Norris}, {da Costa}, {Milone}, {Ivans}, {Gonzalez}, {Fulbright}, {Hilker},
  {Piotto}, {Zoccali}, \& {Stetson}}]{marino+2011b}
{Marino}, A.~F., {Sneden}, C., {Kraft}, R.~P., {et~al.} 2011, \aap, 532, A8

\bibitem[{{Marino} {et~al.}(2008){Marino}, {Villanova}, {Piotto}, {Milone},
  {Momany}, {Bedin}, \& {Medling}}]{marino+2008}
{Marino}, A.~F., {Villanova}, S., {Piotto}, G., {et~al.} 2008, \aap, 490, 625

\bibitem[{{Matrozis} {et~al.}(2013){Matrozis}, {Ryde}, \&
  {Dupree}}]{matrozis+2013}
{Matrozis}, E., {Ryde}, N., \& {Dupree}, A.~K. 2013, \aap, 559, A115

\bibitem[{{Matteucci} \& {Brocato}(1990)}]{matteucci+brocato1990}
{Matteucci}, F. \& {Brocato}, E. 1990, \apj, 365, 539

\bibitem[{{McWilliam}(1997)}]{mcwilliam1997}
{McWilliam}, A. 1997, \araa, 35, 503

\bibitem[{McWilliam {et~al.}(1995)McWilliam, Preston, Sneden, \&
  Searle}]{mcwilliam+95}
McWilliam, A., Preston, G.~W., Sneden, C., \& Searle, L. 1995, AJ, 109, 2757

\bibitem[{{McWilliam} {et~al.}(2013){McWilliam}, {Wallerstein}, \&
  {Mottini}}]{mcwilliam+2013}
{McWilliam}, A., {Wallerstein}, G., \& {Mottini}, M. 2013, \apj, 778, 149

\bibitem[{{Mishenina} {et~al.}(2006){Mishenina}, {Bienaym{\'e}}, {Gorbaneva},
  {Charbonnel}, {Soubiran}, {Korotin}, \& {Kovtyukh}}]{mishenina+2006}
{Mishenina}, T.~V., {Bienaym{\'e}}, O., {Gorbaneva}, T.~I., {et~al.} 2006,
  \aap, 456, 1109

\bibitem[{{Mishenina} {et~al.}(2007){Mishenina}, {Gorbaneva}, {Bienaym{\'e}},
  {Soubiran}, {Kovtyukh}, \& {Orlova}}]{mishenina+2007}
{Mishenina}, T.~V., {Gorbaneva}, T.~I., {Bienaym{\'e}}, O., {et~al.} 2007,
  Astronomy Reports, 51, 382

\bibitem[{{Monaco} {et~al.}(2012){Monaco}, {Villanova}, {Bonifacio}, {Caffau},
  {Geisler}, {Marconi}, {Momany}, \& {Ludwig}}]{monaco+2012}
{Monaco}, L., {Villanova}, S., {Bonifacio}, P., {et~al.} 2012, \aap, 539, A157

\bibitem[{{Nissen} {et~al.}(2007){Nissen}, {Akerman}, {Asplund}, {Fabbian},
  {Kerber}, {Kaufl}, \& {Pettini}}]{nissen+2007}
{Nissen}, P.~E., {Akerman}, C., {Asplund}, M., {et~al.} 2007, \aap, 469, 319

\bibitem[{{Nissen} {et~al.}(2004){Nissen}, {Chen}, {Asplund}, \&
  {Pettini}}]{nissen+2004}
{Nissen}, P.~E., {Chen}, Y.~Q., {Asplund}, M., \& {Pettini}, M. 2004, \aap,
  415, 993

\bibitem[{{Savage} \& {Sembach}(1996)}]{savage+sembach1996}
{Savage}, B.~D. \& {Sembach}, K.~R. 1996, \araa, 34, 279

\bibitem[{{Sbordone} {et~al.}(2009){Sbordone}, {Limongi}, {Chieffi}, {Caffau},
  {Ludwig}, \& {Bonifacio}}]{sbordone+2009}
{Sbordone}, L., {Limongi}, M., {Chieffi}, A., {et~al.} 2009, \aap, 503, 121

\bibitem[{Shetrone {et~al.}(2003)Shetrone, Venn, Tolstoy, Primas, Hill, \&
  Kaufer}]{shetrone+2003}
Shetrone, M., Venn, K.~A., Tolstoy, E., {et~al.} 2003, AJ, 125, 684

\bibitem[{Shetrone {et~al.}(2001)Shetrone, C\^{o}t\'e, \&
  Sargent}]{shetrone+2001}
Shetrone, M.~D., C\^{o}t\'e, P., \& Sargent, W. L.~W. 2001, ApJ, 548, 592

\bibitem[{{Spite} {et~al.}(2011){Spite}, {Caffau}, {Andrievsky}, {Korotin},
  {Depagne}, {Spite}, {Bonifacio}, {Ludwig}, {Cayrel}, {Fran{\c c}ois}, {Hill},
  {Plez}, {Andersen}, {Barbuy}, {Beers}, {Molaro}, {Nordstr{\"o}m}, \&
  {Primas}}]{spite+2011}
{Spite}, M., {Caffau}, E., {Andrievsky}, S.~M., {et~al.} 2011, \aap, 528, A9

\bibitem[{{Takeda} {et~al.}(2005){Takeda}, {Hashimoto}, {Taguchi}, {Yoshioka},
  {Takada-Hidai}, {Saito}, \& {Honda}}]{takeda+2005}
{Takeda}, Y., {Hashimoto}, O., {Taguchi}, H., {et~al.} 2005, \pasj, 57, 751

\bibitem[{{Takeda} \& {Takada-Hidai}(2011)}]{takeda+takada-hidai2011}
{Takeda}, Y. \& {Takada-Hidai}, M. 2011, \pasj, 63, 537

\bibitem[{Tinsley(1979)}]{tinsley1979}
Tinsley, B.~M. 1979, ApJ, 229, 1046

\bibitem[{{Tolstoy} {et~al.}(2009){Tolstoy}, {Hill}, \& {Tosi}}]{tolstoy+09}
{Tolstoy}, E., {Hill}, V., \& {Tosi}, M. 2009, ARA\&A, 47, 371

\bibitem[{{Ueda} {et~al.}(2005){Ueda}, {Mitsuda}, {Murakami}, \&
  {Matsushita}}]{ueda+2005}
{Ueda}, Y., {Mitsuda}, K., {Murakami}, H., \& {Matsushita}, K. 2005, \apj, 620,
  274

\bibitem[{{Valenti} \& {Piskunov}(1996)}]{valenti+piskunov96}
{Valenti}, J.~A. \& {Piskunov}, N. 1996, \aaps, 118, 595

\bibitem[{Venn {et~al.}(2004)Venn, Irwin, Shetrone, Tout, Hill, \&
  Tolstoy}]{venn+2004}
Venn, K.~A., Irwin, M., Shetrone, M.~D., {et~al.} 2004, AJ, 128, 1177

\bibitem[{{Zerne} {et~al.}(1997){Zerne}, {Caiyan}, {Berzinsh}, \&
  {Svanberg}}]{ZCBS}
{Zerne}, R., {Caiyan}, L., {Berzinsh}, U., \& {Svanberg}, S. 1997, \physscr,
  56, 459, (ZCBS)

\end{thebibliography}


\end{document}